\documentclass[noshowpacs,
amsmath,
prl,
twocolumn,
superscriptaddress,
8pt,
aps]{revtex4-2}

\usepackage{setspace}
\usepackage{amsmath}
\usepackage{multirow}
\usepackage{graphicx}

\usepackage{verbatim}
\usepackage{amsfonts}
\usepackage{amssymb}
\usepackage{epstopdf}
\usepackage{dcolumn}
\usepackage{xcolor}
\usepackage{verbatim}
\usepackage{siunitx}
\usepackage[version=4]{mhchem}
\usepackage{textcomp}
\usepackage[left]{lineno}
\DeclareGraphicsExtensions{.pdf,.eps,.png,.jpg,.mps}

\begin{document}
\title{Raman-induced dynamics of ultrafast microresonator solitons}

\author{Binbin Nie}
\affiliation{State Key Laboratory for Artificial Microstructure and Mesoscopic Physics and Frontiers Science Center for Nano-optoelectronics, School of Physics, Peking University, Beijing 100871, China}

\author{Yuanlei Wang}
\affiliation{State Key Laboratory for Artificial Microstructure and Mesoscopic Physics and Frontiers Science Center for Nano-optoelectronics, School of Physics, Peking University, Beijing 100871, China}

\author{Du Qian}
\affiliation{State Key Laboratory for Artificial Microstructure and Mesoscopic Physics and Frontiers Science Center for Nano-optoelectronics, School of Physics, Peking University, Beijing 100871, China}

\author{Yiwen Yang}
\affiliation{State Key Laboratory for Artificial Microstructure and Mesoscopic Physics and Frontiers Science Center for Nano-optoelectronics, School of Physics, Peking University, Beijing 100871, China}

\author{Haoyang Luo}
\affiliation{State Key Laboratory for Artificial Microstructure and Mesoscopic Physics and Frontiers Science Center for Nano-optoelectronics, School of Physics, Peking University, Beijing 100871, China}

\author{Junqi Wang}
\affiliation{State Key Laboratory for Artificial Microstructure and Mesoscopic Physics and Frontiers Science Center for Nano-optoelectronics, School of Physics, Peking University, Beijing 100871, China}

\author{Yun-Feng Xiao}
\affiliation{State Key Laboratory for Artificial Microstructure and Mesoscopic Physics and Frontiers Science Center for Nano-optoelectronics, School of Physics, Peking University, Beijing 100871, China}
\affiliation{Peking University Yangtze Delta Institute of Optoelectronics, Nantong, Jiangsu 226010, China}
\affiliation{Collaborative Innovation Center of Extreme Optics, Shanxi University, Taiyuan 030006, China}

\author{Qihuang Gong}
\affiliation{State Key Laboratory for Artificial Microstructure and Mesoscopic Physics and Frontiers Science Center for Nano-optoelectronics, School of Physics, Peking University, Beijing 100871, China}
\affiliation{Peking University Yangtze Delta Institute of Optoelectronics, Nantong, Jiangsu 226010, China}
\affiliation{Collaborative Innovation Center of Extreme Optics, Shanxi University, Taiyuan 030006, China}

\author{Qi-Fan Yang}
\email[Contact author: ]{leonardoyoung@pku.edu.cn}
\affiliation{State Key Laboratory for Artificial Microstructure and Mesoscopic Physics and Frontiers Science Center for Nano-optoelectronics, School of Physics, Peking University, Beijing 100871, China}
\affiliation{Peking University Yangtze Delta Institute of Optoelectronics, Nantong, Jiangsu 226010, China}
\affiliation{Collaborative Innovation Center of Extreme Optics, Shanxi University, Taiyuan 030006, China}

\begin{abstract}
Soliton microcombs are evolving towards octave-spanning for $f$–$2f$ self-referencing and expanding applications in spectroscopy and timekeeping. As spectra broaden and pulses shorten, the Raman-induced soliton self-frequency shift (SSFS) becomes a principal limitation: it reduces pump–to–comb conversion efficiency, constrains achievable span, and can, in extremes, preclude stationary operation. We develop a complementary theory of SSFS in microresonators that remains valid when the soliton duration $\tau_s$ is shorter than the Raman response timescale. The theory predicts a reduced dependence of the SSFS on $\tau_s$ which also expands the soliton existence range. Such predictions are validated by numerical simulations and by experiments on Si$_3$N$_4$ microresonators. Our results provide practical guidelines for engineering efficient and broadband soliton microcombs.
\end{abstract}

\maketitle

\begin{figure*}[!t]
\centering
    \includegraphics[scale = 1.0]{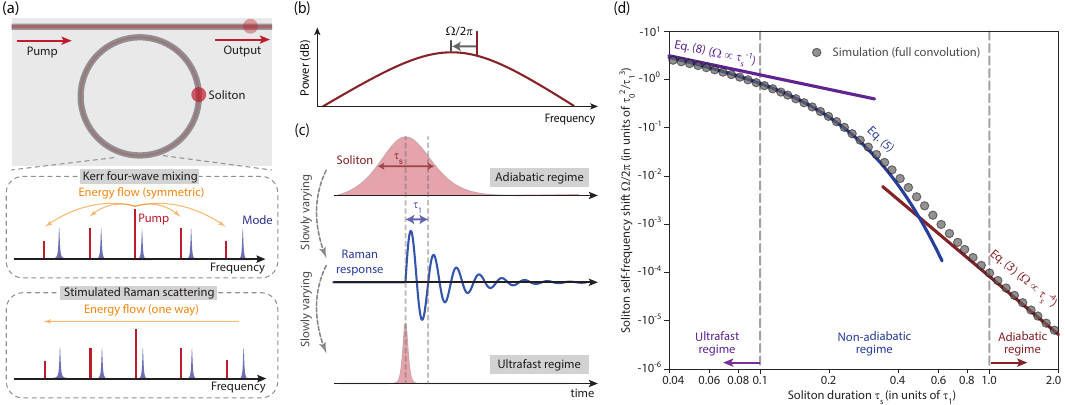}
    \caption{
    (a) Schematic of a microresonator soliton and the intracavity nonlinear processes.
    (b) Raman-induced SSFS \((\Omega/2\pi)\) in a soliton spectrum, manifested as a red-shift of the soliton’s spectral center relative to the pump.
    (c) Temporal illustration of soliton pulses and the material Raman response. 
    (d) Simulated SSFS (gray dots) and analytical predictions (solid lines) versus soliton duration \(\tau_s\). Lorentzian Raman response [Eq.~\eqref{hR}] with $\tau_2/\tau_1 = 1$ is used in the plot. 
    }
    \label{figure1}
\end{figure*}

\textit{Introduction--} Continuous-wave pumped optical microresonators support dissipative Kerr solitons—self-sustained pulses maintained by a double balance of dispersion with nonlinearity and gain with loss—thereby enabling chip-scale optical frequency combs~\cite{Kippenberg2018}. In the frequency domain, the soliton train appears as an evenly spaced, highly coherent set of lines that underpins applications from telecommunications~\cite{marin2017microresonator} to precision spectroscopy~\cite{suh2016microresonator,dutt2018chip,yang2019vernier} and timekeeping~\cite{newman2019architecture,wu2025vernier}. Energy transfer from the pump to the comb lines is mediated predominantly by Kerr four-wave mixing and is symmetric about the pump~\cite{kippenberg2004kerr,del2007optical} [Fig.~\ref{figure1}(a)]. By contrast, the material Raman response drives a unidirectional energy flow toward lower optical frequencies [Fig.~\ref{figure1}(a)], producing a Raman-induced soliton self-frequency shift (SSFS) that manifests as a red shift of the soliton spectral center relative to the pump~\cite{yi2016theory,karpov2016raman} [Fig.~\ref{figure1}(b)]. While the SSFS can be leveraged to tune the comb spacing via group-velocity dispersion~\cite{yang2016spatial,yi2017single}, it also induces an exponential reduction of pump-to-comb conversion efficiency with increasing spectral span~\cite{yi2016theory}, and excessive SSFS can disrupt steady soliton operation~\cite{wang2018stimulated,chen2018experimental}. Consequently, a quantitative description of SSFS is essential for designing broadband soliton microcombs \cite{song2024octave,pfeiffer2017octave,wang2024octave,zang2024foundry,liu2021aluminum,li2017stably}, which are critical for $f$–$2f$ self-referencing \cite{telle1999carrier}.

The amount of Raman-induced SSFS is dependent on the material and soliton properties \cite{gordon1986theory,karpov2016raman,yi2016theory}. Prior studies have focused on relatively long-duration solitons with pulse duration $\tau_s \gg \tau_1$, where $\tau_1$ denotes the characteristic Raman response time of the medium~\cite{gordon1986theory,karpov2016raman,yi2016theory}. In this adiabatic regime [Fig.~\ref{figure1}(c)], the soliton envelope varies slowly compared to the Raman response, and the SSFS follows Gordon’s scaling $\propto \tau_s^{-4}$~\cite{gordon1986theory,yi2016theory}. Recently, the spectral span of soliton microcombs has been significantly improved due to engineered dispersion and efficient pumping schemes~\cite{song2024octave,helgason2023surpassing,zhu2025ultra,yang2024efficient}. They correspond to ultrafast solitons with $\tau_s \ll \tau_1$, such that the Raman response evolves slowly relative to the soliton envelope. This inversion suggests a substantially modified SSFS scaling, yet a quantitative theory for this regime has been lacking.

In this work, we develop a complementary theory of Raman-induced SSFS for microresonator solitons with a wide range of pulse durations and validate it through numerical simulations and experiments. The theory predicts a progressive weakening of the SSFS dependence on $\tau_s$ as the pulse shortens, resulting in markedly reduced SSFS for ultrashort solitons relative to adiabatic expectations. It also expands the soliton existence range and enables broader, more efficient microcomb operation.

\textit{Theory--} We begin with the Raman-augmented Lugiato–Lefever equation (LLE) for the intracavity field in a Kerr microresonator~\cite{agrawal2007nonlinear,lugiato1987spatial,lv2025broadband,milian2015solitons}:
\begin{equation}
\begin{aligned}
\frac{\partial A(T,t)}{\partial T}=&-\left(\frac{\kappa}{2}+i\delta\omega\right)A
-\frac{ic\beta_2}{2n_0}\frac{\partial^2 A}{\partial t^2}
+\sqrt{\kappa_{\mathrm{ext}}P_{\mathrm{in}}}\\
&+igA\int R(t^\prime) \lvert A(t-t^\prime)\rvert^2 \mathrm{d}t^\prime,
\end{aligned}
\label{Main:lle}
\end{equation}
where $T$ is the slow evolution time, $t$ is the fast retarded time in a frame co-rotating at the free-spectral-range (FSR) rate, and $A(T,t)$ is the slowly varying intracavity envelope normalized so that $\lvert A\rvert^2$ equals the intracavity energy. The total decay rate is $\kappa=\kappa_{\mathrm{int}}+\kappa_{\mathrm{ext}}$, with $\kappa_{\mathrm{int}}$ the intrinsic decay rate and $\kappa_{\mathrm{ext}}$ the external coupling rates. $\delta\omega$ is the laser–cavity detuning, $\beta_2$ the group-velocity dispersion, $c$ the speed of light, and $P_{\mathrm{in}}$ the pump power. The Kerr nonlinearity coefficient is $g=\omega_0 n_2 D_1/(2\pi n_0 A_{\mathrm{eff}})$, where $n_0$ and $n_2$ are the linear and nonlinear refractive indices, $D_1/2\pi$ is the FSR, and $A_{\mathrm{eff}}$ is the effective mode area. The material nonlinear response is~\cite{agrawal2007nonlinear,lin2007nonlinear}
\begin{equation}
R(t)=(1-f_R)\delta(t)+f_Rh_R(t),
\label{main:nonlinear_response}
\end{equation}
combining an almost instantaneous electronic (Kerr) contribution and a delayed vibrational (Raman) contribution. Here $f_R$ is the fractional Raman weight, $\delta(t)$ is the Dirac delta, and $h_R(t)$ is the Raman response function (normalized so that $\int_{-\infty}^{+\infty} h_R(t)\mathrm{d}t=1$).

Treating resonator dissipation, external pumping, and the delayed Raman response in Eq.~\eqref{Main:lle} as weak perturbations, an analytic expression for the SSFS can be obtained via the perturbed Lagrangian method~\cite{matsko2013timing,yi2016theory}. In the adiabatic regime ($\tau_s \gg \tau_1$), it gives
\begin{equation}
\Omega(\tau_s)=-\frac{8\tau_o^2}{15}\frac{\tau_{\mathrm A}}{\tau_s^{4}},
\label{adiabatic}
\end{equation}
where $\Omega$ is the SSFS in angular frequency, $\tau_o=\sqrt{c\lvert\beta_2\rvert/(\kappa n_0)}$ is a characteristic cavity timescale, and $\tau_{\mathrm A}=f_R\int_{-\infty}^{+\infty}\tau h_R(\tau) \mathrm{d}\tau$ is the Raman shock time~\cite{agrawal2007nonlinear}. In this adiabatic limit, the molecular vibration is not temporally resolved by the much longer soliton; the Raman convolution effectively reduces to the constant $\tau_{\mathrm A}$, yielding the fixed scaling $\Omega\propto\tau_s^{-4}$ [Fig.~\ref{figure1}(d), red].

The adiabatic approximation fails when $\tau_s \lesssim \tau_1$, i.e., in the non-adiabatic regime. In this regime, the soliton spectrum is sufficiently broad that it varies little across the Raman peak bandwidth. Under this assumption, the Raman convolution reduces to (see Supplemental Material)
\begin{equation}
    \int h_R(t^\prime) \left| A(t-t^\prime)\right|^2 \mathrm{d}t^\prime \approx 2\pi \widetilde{S}\left(\frac{2\pi}{\tau_1}\right) h_R(t) + \left|A(t)\right|^2
\end{equation}
where $\widetilde{S}(\omega) = \frac{1}{2\pi}\int \left|A(t)\right|^2e^{i\omega t} \mathrm{d}t$ is the Fourier transform of the soliton intensity. Employing this approximation and the Lagrangian perturbation method (see Supplemental Material), we obtain the analytic SSFS expression 
\begin{equation}
\Omega(\tau_s)=-2\tau_o^2\frac{\pi^2\tau_s/\tau_1}{\sinh\left(\pi^2\tau_s/\tau_1\right)}\frac{\theta_{\mathrm U}(\tau_s)}{\tau_s^3},
\label{uf_trans}
\end{equation}
where $\theta_{\mathrm U}(\tau_s)$—the counterpart of the Raman shock time $\tau_{\mathrm A}$—is a duration-dependent functional of the Raman response, defined as
\begin{equation}
\theta_{\mathrm U}(\tau_s)=f_R\int_{-\infty}^{+\infty} h_R(\tau)
\operatorname{sech}^2\left(\frac{\tau}{\tau_s}\right)
\tanh\left(\frac{\tau}{\tau_s}\right)\mathrm{d}\tau.
\label{thetaU}
\end{equation}
Crucially, the ratio $\tau_s/\tau_1$ enters Eq.~\eqref{uf_trans} through the factor $x/\sinh x$ with $x=\pi^2\tau_s/\tau_1$, leading to an exponential suppression of the SSFS and a progressively weaker dependence on $\tau_s$ as the pulse shortens [Fig.~\ref{figure1}(d), blue]. Intuitively, in the time domain, the delayed Raman polarization cannot fully develop within an ultrashort pulse, thereby diminishing the net SSFS.

Although our framework is, in principle, independent of the specific Raman response, an explicit model is required for closed-form analysis and quantitative predictions. We adopt the standard damped-oscillator model, which yields a Lorentzian Raman gain in the frequency domain~\cite{agrawal2007nonlinear,lin2007nonlinear}. The corresponding temporal response is
\begin{equation}
h_R(t)=\frac{\left(\tau_1/2\pi\right)^2+\tau_2^2}{\left(\tau_1/2\pi\right)\tau_2^2}
\sin\left(2\pi\frac{t}{\tau_1}\right)\exp\left(-\frac{t}{\tau_2}\right)\mathrm{H}(t),
\label{hR}
\end{equation}
where $\tau_1$ and $\tau_2$ are the Raman vibrational period and lifetime, respectively, and $\mathrm{H}(t)$ is the Heaviside step function enforcing causality ($h_R(t)=0$ for $t<0$).

In the ultrafast regime ($\tau_s\ll\tau_1$), substituting Eq.~\eqref{hR} into Eq.~\eqref{thetaU} yields
$\theta_{\mathrm U}(\tau_s)\approx 2f_R\pi^2\tau_s^2/\tau_1^{2}$. Using $x=\pi^2\tau_s/\tau_1$ and $\lim_{x\to 0}x/\sinh x=1$ in Eq.~\eqref{uf_trans} then gives
\begin{equation}
\Omega(\tau_s)=-4\pi^2\tau_o^2\frac{f_R}{\tau_1^{2}}\frac{1}{\tau_s},
\label{uf_eventual}
\end{equation}
which exhibits the characteristic $\tau_s^{-1}$ scaling [Fig.~\ref{figure1}(d), purple]. To validate these analytical results, we perform benchmark simulations by numerically integrating the LLE with full convolution [Eq.~\eqref{Main:lle}]. The analytical predictions show excellent agreement with the simulation results (gray circles), thereby validating our theory.

\begin{figure}[tbp]
\centering
    \includegraphics[scale = 1.0]{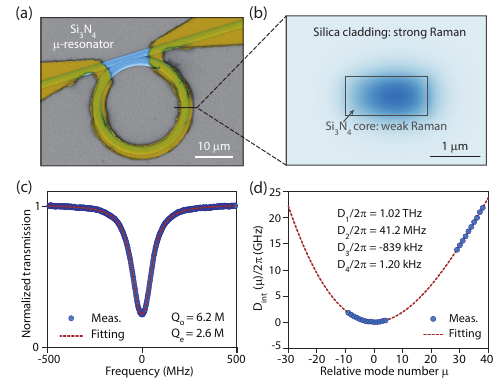}
    \caption{
    (a) False-colored scanning electron microscope image of the Si$_3$N$_4$ microresonator. The Si$_3$N$_4$ structure and metal heater are shown in blue and gold, respectively.     
    (b) Cross-sectional profile of the fundamental transverse-electric (TE$_{00}$) mode.  
    (c) Transmission spectrum of a resonance near 1546 nm. The fitting reveals intrinsic and external-coupling quality factors of $6.2\times10^6$ and $2.6\times10^6$, respectively.
    (d) Measured integrated dispersion $D_\mathrm{int}$ (blue circles) with a polynomial fitting (red dashed line). The dispersion parameters are extracted from the fitting. 
    }
    \label{figure2}
\end{figure}

\begin{figure*}[htbp]
\centering
    \includegraphics[scale = 1.0]{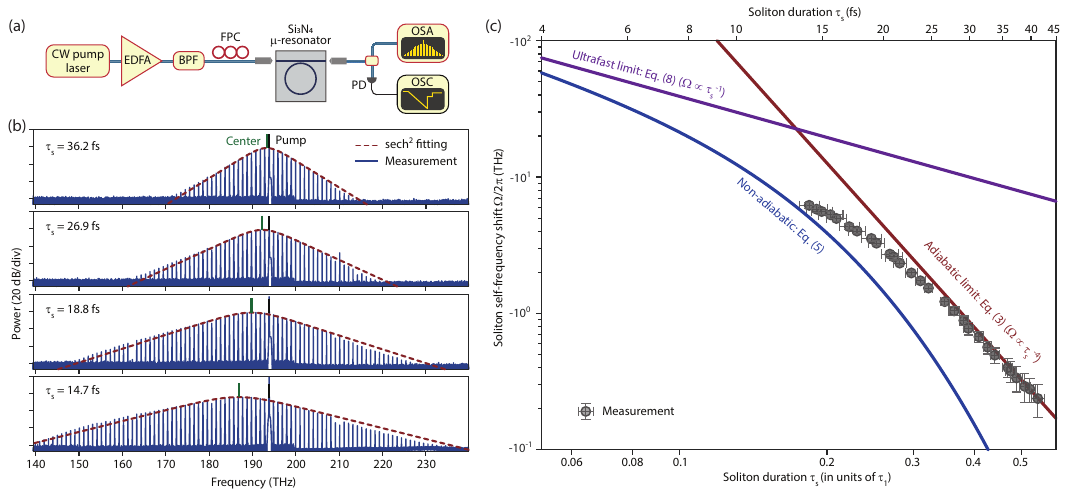}
    \caption{
    (a) Experimental setup. CW: continuous wave; EDFA: erbium-doped fiber amplifier; BPF: band-pass filter; FPC: fiber polarization controller; PD: photodetector; OSC: oscilloscope; OSA: optical spectrum analyzer. 
    (b) Optical spectra of soliton microcombs with different soliton durations. The spectral envelopes are fitted with \(\mathrm{sech}^2\) functions (dashed lines), from which the soliton durations and SSFS values are extracted.
    (c) Measured SSFS (gray circles) and analytical predictions (solid lines) versus soliton duration \(\tau_s\). As \(\tau_s\) decreases, the measurements deviate from the scaling law of adiabatic limit (red, Eq.~\eqref{adiabatic}) and converge toward the prediction of non-adiabatic theory (blue, Eq.~\eqref{uf_trans}). 
    The error bars indicate the 95\% confidence interval from nonlinear least-squares fitting. 
    }
    \label{figure3}
\end{figure*}

\textit{Experiment--} We perform proof-of-concept experiments to validate the theory using a Si$_3$N$_4$ microresonator fabricated by subtractive processing~\cite{wang2025compact} [Fig.~\ref{figure2}(a)]. The device has radius 22.3~\textmu{m}, FSR $\approx 1~\mathrm{THz}$, and supports a fundamental transverse-electric (TE$_{00}$) mode whose simulated profile is shown in Fig.~\ref{figure2}(b). Because the Si$_3$N$_4$ core has very weak Raman response, the Raman response of the microresonator is dominated by the silica cladding (with $f_R^{\mathrm{silica}}=0.18$, $\tau_1/2\pi=12.2~\mathrm{fs}$, and $\tau_2=32~\mathrm{fs}$)~\cite{agrawal2007nonlinear,zheng2025silicon}, yielding an effective Raman fraction $f_R=0.0217$ from finite-element eigenmode simulations [Fig.~\ref{figure2}(b)]. A resonance near $1546~\mathrm{nm}$ exhibits intrinsic and coupling quality factors of $6.2\times10^{6}$ and $2.6\times10^{6}$, respectively [Fig.~\ref{figure2}(c)]. The integrated dispersion, defined as $D{_\mathrm{int}}(\mu)=\omega_\mu-\omega_0-D_1\mu=\sum_{n\ge2}D_n\mu^n/n!$ (with integer mode index $\mu$ relative to the pump), extracted from the measured resonance frequencies~\cite{li2012sideband}, indicates anomalous group-velocity dispersion ($D_2 = -cD_1^2\beta_2/n_o$) with only minor higher-order contributions [Fig.~\ref{figure2}(d)]. 

In the experiment, the Si$_3$N$_4$ microresonator is pumped by an amplified continuous-wave laser to generate solitons [Fig.~\ref{figure3}(a)]. The optical spectra are measured using two optical spectrum analyzers (OSA, Yokogawa AQ6374, AQ6376). The soliton duration $\tau_s$ and Raman-induced SSFS $\Omega/2\pi$ (in Hz) are extracted by fitting a $\mathrm{sech}^2$ envelope to the measured spectra (pump line excluded) and determining the spectral center relative to the pump [Fig.~\ref{figure3}(b)]. Small discrepancies between the sech${}^2$ fit and the measured spectra are attributed to wavelength-dependent transmission along the path from the microresonator to the OSA, including microresonator–waveguide coupling and fiber components (e.g., couplers). We restrict the fitting window to 184–198~THz, where these wavelength-dependent effects can be calibrated using a continuously tunable laser (Toptica CTL 1550) (see Supplementary Materials). Within the accessible bandwidths, $D_3\mu^3/6\ll D_2\mu^2/2$, so frequency shifts from asymmetric dispersion are negligible. Self-steepening~\cite{agrawal2007nonlinear,bao2015soliton} is likewise negligible for our operating conditions, as validated using numerical simulations (see Supplementary Materials).

Figure~\ref{figure3}(c) compares the measured SSFS with analytic predictions. For longer-duration solitons, the data follow the $\tau_s^{-4}$ scaling in excellent agreement with the adiabatic theory~\cite{yi2016theory} [red line, Eq.~\eqref{adiabatic}], yielding a Raman shock time $\tau_{\mathrm A}\approx 0.73~\mathrm{fs}$, enhanced by the Boson-peak in the silica Raman spectrum~\cite{hu2024theory,lin2006raman}. As $\tau_s$ decreases, the soliton envelope is no longer slowly varying on the silica Raman timescale and the SSFS progressively departs from $\tau_s^{-4}$ scaling, approaching the non-adiabatic prediction [blue line, Eq.~\eqref{uf_trans}]. Although the available pump power and telecom-band carrier preclude reaching the strict ultrafast limit (well below 10 fs), the measurements clearly capture the adiabatic–to–non-adiabatic transition, providing strong evidence for the modified scaling law. The actual ultrafast regime can be potentially accessed in materials with longer Raman vibration periods (e.g., $\tau_1\approx132~\mathrm{fs}$ for the dominant $\mathrm{A}(\mathrm{TO})_1$ mode in LiNbO$_3$~\cite{bache2012review,nie2025soliton}), or using efficient pumping schemes that can result in broader soliton microcombs at given pump power~\cite{helgason2023surpassing,zhu2025ultra,yang2024efficient}.

\begin{figure}[!t]
\centering
    \includegraphics[scale = 1.0]{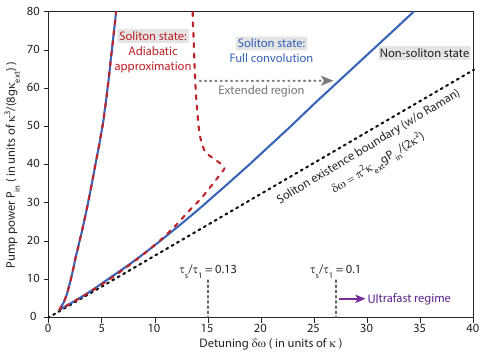}
    \caption{Soliton existence range predicted by the adiabatic approximation (red dashed line), the full convolution model (blue solid line), and the model without Raman (black dashed line). The parameters used in simulating the diagram are: $\tau_1/2\pi = 12.2~\mathrm{fs},\tau_2=32~\mathrm{fs},f_R=0.05,D_1/2\pi = 500~\mathrm{GHz},\kappa/2\pi= 193~\mathrm{MHz}, \kappa_\mathrm{ext}/\kappa=0.5 ,D_2/2\pi = 5~\mathrm{MHz}$. 
    }
    \label{figure4}
\end{figure}
\textit{Soliton existence range--} In the absence of Raman-induced SSFS, the largest detuning ($\delta\omega_\mathrm{max}$) attainable at a given input power ($P_\mathrm{in}$) is $\delta\omega_\mathrm{max}=\pi^{2}\kappa_\mathrm{ext}gP_\mathrm{in}/(2\kappa^{2})$ \cite{herr2014temporal} [Fig.~\ref{figure4}]. At maximum detuning, a soliton microcomb attains its maximum spectral span. When SSFS is included, the maximum accessible detuning is reduced, scaling as $\mathrm{sech}^2\left(\pi\Omega\tau_s/2 \right)$ relative to the non-Raman case~\cite{yi2016theory}. In the extreme of sufficiently large SSFS, the resulting dynamical instability precludes stationary solitons irrespective of pump power \cite{wang2018stimulated}. Because the adiabatic scaling ($\Omega\propto\tau_s^{-4}$) overestimates the SSFS for short solitons, we re-evaluate the soliton existence range using the full convolution simulation [Eq.~\eqref{Main:lle}]. As shown in Fig.~\ref{figure4}, for our parameters, the adiabatic approximation (Eq.~(20) in Supplementary Material) predicts a notably narrower existence range than the non-Raman case and clamps the upper detuning to $\simeq 15\kappa$. By contrast, the full convolution simulation [Eq.~\eqref{Main:lle}] yields an existence range much closer to the non-Raman case, with the upper detuning continuing to increase with pump power. These results indicate that, even in the presence of Raman effects, soliton microcombs are still possible to reach ultrabroad spectra.

\textit{Discussion--} We develop a complementary theory of Raman-induced SSFS for microresonator solitons. The theory can be applied to different materials, and in the presence of multiple Raman modes, the response $h_R(t)$ can be treated as a linear superposition of the contributing modes. The accurate prediction of SSFS is critical for designing efficient, ultrabroadband soliton microcombs, with direct applications in miniature optical clocks~\cite{newman2019architecture,wu2025vernier}, astronomical spectrographs~\cite{suh2019searching,obrzud2019microphotonic}, and optical frequency synthesis~\cite{spencer2018optical}. Finally, the theory also indicates the possibility of realizing few-cycle pulses in microresonators even in the presence of Raman nonlinearity, particularly given recent advances in highly efficient soliton pumping schemes~\cite{yang2024efficient,zhu2025ultra,helgason2023surpassing}, as well as methods that use dispersion-induced spectral recoiling \cite{milian2015solitons,brasch2016photonic,yi2017single} to balance Raman-induced SSFS. Similar to the modified Raman dynamics presented in this work, the increased span of soliton microcombs could motivate modifications of the LLE appropriate to the ultrafast limit.

\medskip
This work was supported by National Key R\&D Plan of China (Grant No. 2023YFB2806702), National Natural Science Foundation of China (12293050), and the High-performance Computing Platform of Peking University. The fabrication in this work was supported by the Peking University Nano-Optoelectronic Fabrication Center, Micro/nano Fabrication Laboratory of Synergetic Extreme Condition User Facility (SECUF), Songshan Lake Materials Laboratory, and the Advanced Photonics Integrated Center of Peking University.

\medskip
B. N. and Y. W. contributed equally to this work.

\bibliography{ref.bib}

\end{document}


\title{Supplementary information: Raman-induced dynamics of ultrafast microresonator solitons}

\author{Binbin Nie$^{1*}$, Yuanlei Wang$^{1*}$, Du Qian$^{1}$, Yiwen Yang$^{1}$, Haoyang Luo$^{1}$, Junqi Wang$^{1}$, Yun-Feng Xiao$^{1,2,3}$, Qihuang Gong$^{1,2,3}$ and Qi-Fan Yang$^{1,2,3,\dagger}$\\
$^1$State Key Laboratory for Artificial Microstructure and Mesoscopic Physics and Frontiers Science Center for Nano-optoelectronics, School of Physics, Peking University, Beijing 100871, China\\
$^2$Peking University Yangtze Delta Institute of Optoelectronics, Nantong, Jiangsu 226010, China\\
$^3$Collaborative Innovation Center of Extreme Optics, Shanxi University, Taiyuan 030006, China\\
$^{*}$These authors contributed equally to this work.\\
$^{\dagger}$Corresponding author: leonardoyoung@pku.edu.cn}

\maketitle

\section{Theoretical analysis}
\subsection{Lagrangian perturbation formalism}
The dynamics of dissipative Kerr soliton in microresonators is governed by the Lugiato–Lefever equation (LLE), which is given by [Eq.~(1)] in the main text]:
\begin{equation}
    \frac{\partial A(T, t)}{\partial T}=-\left(\frac{\kappa}{2} + i \delta \omega\right) A- \frac{ic\beta_2}{2n_o} \frac{\partial^2 A}{\partial t^2}+\sqrt{{\kappa_{\mathrm{ext}} P_{\mathrm{in}}}}+ i(1-f_R)g\left|A\right|^2A
    +i g f_R A  \int
    h_R(t^\prime)\left|A(t-t^\prime)\right|^2 \mathrm{d}t^\prime. 
\label{SI:lle}
\end{equation}
The physical quantities involved in Eq.~\eqref{SI:lle} are defined in the main text. We employ the Lagrangian perturbation method~\cite{matsko2013timing, yi2016theory} to analytically derive the Raman-induced soliton self-frequency shift (SSFS) in the steady state. In the absence of dissipation, pumping, and Raman response, the microresonator is a conservative system, whose Lagrangian density is given by~\cite{matsko2013timing, yi2016theory}:
\begin{equation}
\mathcal{L}=\frac{1}{2}\left(A^* \frac{\partial A}{\partial T}-A \frac{\partial A^*}{\partial T}\right)-\frac{ic \beta_2}{2 n_o}\left|\frac{\partial A}{\partial t}\right|^2-\frac{i}{2} g|A|^4+i\delta \omega|A|^2. 
\label{SI:LagDensity}
\end{equation}
The dissipation, pumping, and Raman response in Eq.~\eqref{SI:lle} are treated as perturbations, collected into
\begin{equation}
\mathcal{R} = -\frac{\kappa}{2}A + \sqrt{{\kappa_{\mathrm{ext}} P_{\mathrm{in}}}}+igf_RA\int
\left[h_R(t^\prime)-\delta(t^\prime)\right]\left|A(t-t^\prime)\right|^2 \mathrm{d}t^\prime. 
\label{SI:perturbation}
\end{equation}
The LLE [Eq.~\eqref{SI:lle}] can be recovered from the variational relation $\delta\mathcal{L}/\delta A^* = \mathcal{R}$~\cite{hasegawa2002soliton}. 

As an ansatz, the slowly varying envelope of a microresonator soliton with a frequency shift is taken as
\begin{equation}A=B \operatorname{sech}\left[(t-t_o)/\tau_s\right] e^{-i \Omega\left(t-t_o\right)+i \varphi}.
\label{SI:ansatz}
\end{equation}
The temporal profile of the soliton ansatz is presented in Fig.~\ref{SIFig:ultrafastregime}(a). When the perturbation is absent, i.e. $\mathcal{R}=0$, the soliton ansatz [Eq.~\eqref{SI:ansatz}] is an exact solution of the LLE [Eq.~\eqref{SI:lle}].
In the following analysis, the soliton parameters—amplitude ($B$), temporal position ($t_o$), duration ($\tau_s$), frequency shift ($\Omega$), and phase ($\varphi$)—are treated as functions evolving on the slow time $T$. We assume soliton operation at far-red detuning ($\delta \omega\gg\kappa$), such that the continuous-wave (CW) background is much weaker than the dominant soliton pulse and can be neglected in the ansatz [Eq.~\eqref{SI:ansatz}].

Substituting the soliton ansatz [Eq.~\eqref{SI:ansatz}] into the Lagrangian density [Eq.~\eqref{SI:LagDensity}] and integrating over the fast time $t$ yields the reduced Lagrangian of the microresonator system:
\begin{equation}
L=\int \mathcal{L} \mathrm{d} t=2 i B^2 \tau_s \left(-\frac{c \beta_2}{6 n_o \tau_s^2}-\frac{g B^2}{3}+\delta \omega
-\frac{c \beta_2}{2 n_o} \Omega^2+\Omega \frac{\partial t_o}{\partial T}+\frac{\partial \varphi}{\partial T}\right).
\label{SI:L}
\end{equation}
The equation of motion for the soliton parameters is governed by the perturbed Euler–Lagrange equation~\cite{hasegawa2002soliton,yi2016theory,matsko2013timing}:
\begin{equation}
\frac{\partial L}{\partial r_i}-\frac{\mathrm{d}}{\mathrm{d} T} \frac{\partial L}{\partial \dot{r}_i}=\int \mathrm{d} t\left(\mathcal{R} \frac{\partial A^*}{\partial r_i}-\mathcal{R}^* \frac{\partial A}{\partial r_i}\right), 
\label{eulerlagrange}
\end{equation}
where the coordinates $r_i$ range over the soliton parameters $\{B, t_o,\tau_s,\Omega,\varphi\}$. By substituting the reduced Lagrangian [Eq.~\eqref{SI:L}] and the perturbation term [Eq.~\eqref{SI:perturbation}] into Eq.~\eqref{eulerlagrange}, analytical expressions for the soliton parameters can, in principle, be derived. However, the nonlocal temporal convolution in the perturbation term renders a closed-form treatment analytically intractable and obscures simple physical laws. Therefore, approximations that capture the essence of the interaction are required.

\subsection{Approximations}
\begin{figure*}[tbp]
    \centering
    \includegraphics[scale=1.0]{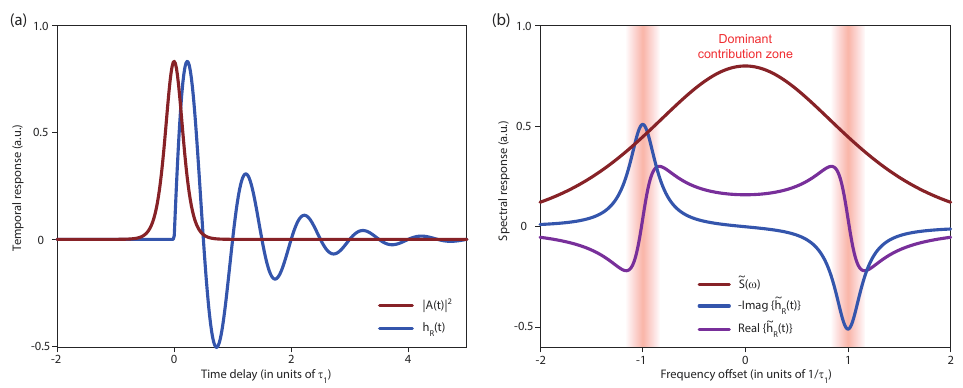}
    \caption{
    (a) Temporal profiles of the soliton intensity \(|A(t)|^2\) and the Raman response function \(h_R(t)\). In the ultrafast regime, the soliton duration is much shorter than the Raman timescale. 
    (b) Spectral profiles of the Fourier transform of the soliton intensity \(\widetilde{S}(\omega)\) and the Raman response \(\widetilde{h_R}(\omega)\). For illustration, the standard Lorentzian Raman model [Eq.~(7) in the main text] is used. Parameters used: $\tau_s/\tau_1 = 0.2, \tau_2/\tau_1 = 1.0$. 
    } 
    \label{SIFig:ultrafastregime}
\end{figure*}

The Raman response is formulated as the convolution of the temporal Raman response function and the intensity of the intracavity optical field [Eq.~\eqref{SI:lle}]. Under the convolution theorem, the time-domain convolution operation is represented by a simple multiplication in the frequency domain:
\begin{equation}
\mathcal{F}_{t\rightarrow \omega}\left\{ \int h_R(t^\prime)|A(t-t^\prime)|^2 \mathrm{d}t^\prime\right\} =2\pi\widetilde{h_R} (\omega) \widetilde{S}(\omega)
\label{SI:conv_theo}
\end{equation}
where $\mathcal{F}_{t\rightarrow \omega} \left\{ \cdot \right\} $ denotes the Fourier transform from time to frequency domain, $\widetilde{h_R}(\omega) = \mathcal{F}_{t\rightarrow \omega} \left\{ h_R(t) \right\}$ is the Raman response in the frequency domain, and $\widetilde{S}(\omega) = \mathcal{F}_{t\rightarrow \omega} \left\{  |A(t)|^2\right\}$ is the Fourier transform of the optical field intensity. By the Wiener–Khinchin theorem, $\widetilde{S}(\omega)$ is equivalently the autocorrelation of the Fourier spectrum of the intracavity field. Substituting the soliton ansatz [Eq.~\eqref{SI:ansatz}], the corresponding $\widetilde{S}(\omega)$ is given by:
\begin{equation}
\widetilde{S}(\omega) = \frac{B^2\tau_s^2}{2} \mathrm{csch} \left( \frac{\pi\tau_s\omega}{2}\right) \omega, 
\label{SI:S}
\end{equation}
which is single-peaked, with a full width at half maximum (FWHM) inversely proportional to the soliton duration $\tau_s$ [Fig.~\ref{SIFig:ultrafastregime}(b), red solid line]. 

In general, the Raman response $h_R(t)$ is a damped oscillation in time with angular frequency set by the lattice vibrational mode and is defined only for $t>0$ by causality [Fig.~\ref{SIFig:ultrafastregime}(a)]. Its Fourier transform $\widetilde{h_R}(\omega)$ gives the spectral Raman response; its imaginary part, $\mathrm{Im}\{\widetilde{h_R}(\omega)\}$, corresponds to the Raman gain spectrum, appearing as a peak centered at the Raman vibrational frequency~\cite{agrawal2007nonlinear}[Fig.~\ref{SIFig:ultrafastregime}(b)]. Leveraging the properties of $\widetilde{h_R}(\omega)$ and $\widetilde{S}(\omega)$ functions, certain approximations can be introduced to simplify the analysis. The adiabatic regime has been treated previously~\cite{yi2016theory,karpov2016raman,gordon1986theory}; here we develop the complementary analysis for the ultrafast regime.

In the ultrafast regime, the soliton duration is shorter than the characteristic timescale of the Raman response [Fig.~\ref{SIFig:ultrafastregime}(a)]. Consequently, in the frequency domain, the bandwidth of $\widetilde{S}(\omega)$ is much broader than that of the Raman gain [Fig.~\ref{SIFig:ultrafastregime}(b)]. The Raman spectrum $\widetilde{h_R}(\omega)$ contributes predominantly near the angular frequency offsets $\pm2\pi/\tau_1$; in this neighborhood, $\widetilde{S}(\omega)$ can be approximated by its leading-order (locally constant) value, i.e. $\widetilde{S}(\omega)\approx\widetilde{S}(2\pi/\tau_1)$. Based on this approximation, the convolution between the soliton intensity and the Raman response [Eq.~\eqref{SI:conv_theo}] can be simplified as
\begin{equation}
\begin{aligned}
         \int h_R(t^\prime)|A(t-t^\prime)|^2 \mathrm{d}t^\prime &=2\pi\mathcal{F}^{-1}_{\omega\rightarrow t}\left\{\widetilde{h_R} (\omega) \widetilde{S}(\omega)\right\}\\
         &=2\pi\mathcal{F}^{-1}_{\omega\rightarrow t}\left\{\widetilde{h_R} (\omega) \widetilde{S}(2\pi/\tau_1) +\widetilde{h_R} (0) \widetilde{S}(\omega)\right\}\\
         &= 2\pi \widetilde{S}(2\pi/\tau_1) h_R(t) + |A(t)|^2.
\end{aligned}
\label{SI:uf_approx}
\end{equation}
Substituting the approximation [Eq.~\eqref{SI:uf_approx}] into Eq.~\eqref{SI:perturbation} yields a simplified expression for the perturbation term in the ultrafast regime:
\begin{equation}
\mathcal{R} \approx -\frac{\kappa}{2}A + \sqrt{{\kappa_{\mathrm{ext}} P_{\mathrm{in}}}} +i {2\pi gf_R\widetilde{S}(2\pi/\tau_1)}h_R\left(t \right)A
\label{SI:RU}
\end{equation}

\subsection{Steady-state solution of soliton parameters}
Substituting the simplified perturbation terms [Eq.~\eqref{SI:RU}] and the soliton ansatz [Eq.~\eqref{SI:ansatz}] into the perturbed Euler–Lagrange equation [Eq.~\eqref{eulerlagrange}] yields the equation of motion for soliton parameters:

\begin{equation}
-\frac{c\beta_2}{6 n_o\tau_s^2}-\frac{2}{3} g B^2+\delta \omega-\frac{c\beta_2}{2n_o}\Omega^2+\Omega\frac{\partial t_o}{\partial T}+\frac{\partial \varphi}{\partial T}
=\pi gf_R \frac{\widetilde{S}(2\pi/\tau_1)}{\tau_s}\int_{-\infty}^{+\infty} h_R\left(\tau\right)\mathrm{sech}^2\left(\tau/\tau_s\right) \mathrm{d}\tau,
\label{SI:eq1}
\end{equation}

\begin{equation}
\frac{c\beta_2}{6 n_o\tau_s^2}-\frac{1}{3} g B^2+\delta \omega-\frac{c\beta_2}{2n_o}\Omega^2+\Omega\frac{\partial t_o}{\partial T}+\frac{\partial \varphi}{\partial T}
=2\pi gf_R \frac{\widetilde{S}(2\pi/\tau_1)}{\tau_s}\int_{-\infty}^{+\infty} h_R\left(\tau\right)\mathrm{sech}^2\left(\tau/\tau_s\right)\mathrm{tanh}\left(\tau/\tau_s\right)\left(\tau/\tau_s\right) \mathrm{d}\tau,
\label{SI:eq2}
\end{equation}

\begin{equation}
\frac{\mathrm{d}}{\mathrm{d}T}(B^2\tau_s\Omega) = -\kappa B^2\tau_s\Omega-2\pi g f_RB^2\frac{\widetilde{S}(2\pi/\tau_1)}{\tau_s}\int_{-\infty}^{+\infty} h_R\left(\tau\right)\mathrm{sech}^2\left(\tau/\tau_s\right)\mathrm{tanh}\left(\tau/\tau_s\right)\mathrm{d}\tau,
\label{SI:eq3}
\end{equation}

\begin{equation}
\frac{\mathrm{d}t_o}{\mathrm{d}T} = \frac{c\beta_2}{n_o}\Omega,
\label{SI:eq4}
\end{equation}

\begin{equation}
\frac{\mathrm{d}}{\mathrm{d}T}(B^2\tau_s) = -\kappa B^2\tau_s + \pi \sqrt{\kappa_\mathrm{ext}P_\mathrm{in}}B \tau_s\cos\varphi\mathrm{sech}\left( \frac{\pi\tau_s\Omega}{2}  \right).
\end{equation}
Combining Eqs.~\eqref{SI:eq1}–\eqref{SI:eq4} and imposing steady-state conditions, the analytical expression for SSFS can be derived:
\begin{equation}
    \Omega(\tau_s) = \frac{c\beta_2}{\kappa n_o} \frac{1}{\tau_s^3}\frac{2 f_R \int_{-\infty}^{+\infty} h_R\left(\tau\right)\mathrm{sech}^2\left(\tau/\tau_s\right)\mathrm{tanh}\left(\tau/\tau_s\right)\mathrm{d}\tau}{\frac{\mathrm{sinh(\pi^2\tau_s/\tau_1)}}{\pi^2\tau_s/\tau_1} + 3f_R \int_{-\infty}^{+\infty} h_R\left(\tau\right)\mathrm{sech}^2\left(\tau/\tau_s\right)\left( 1-2\left(\tau/\tau_s\right)\mathrm{tanh}\left(\tau/\tau_s\right)\right)\mathrm{d}\tau}. 
\end{equation}
Noting that, over the parameter range considered,
\begin{equation} \frac{\mathrm{sinh(\pi^2\tau_s/\tau_1)}}{\pi^2\tau_s/\tau_1} \gg f_R \int_{-\infty}^{+\infty} h_R\left(\tau\right)\mathrm{sech}^2\left(\tau/\tau_s\right)\left( 1-2\left(\tau/\tau_s\right)\mathrm{tanh}\left(\tau/\tau_s\right)\right)\mathrm{d}\tau,
\end{equation}
the SSFS reduces to
\begin{equation}
    \Omega(\tau_s)\approx 2\frac{c\beta_2}{\kappa n_o}\frac{\pi^2\tau_s/\tau_1}{\mathrm{sinh}\left({\pi^2\tau_s/\tau_1}\right)} \frac{f_R \int_{-\infty}^{+\infty} h_R\left(\tau\right)\mathrm{sech}^2\left(\tau/\tau_s\right)\mathrm{tanh}\left(\tau/\tau_s\right)\mathrm{d}\tau}{\tau_s^3},
\end{equation}
which corresponds to Eq.~(5) in the main text.

\section{Simulation}
In Fig.~4 of the main text, under the adiabatic (shock-time) approximation, the Raman convolution reduces to a Raman shock term. The corresponding master equation (LLE) employed in the simulation is given by \cite{yi2016theory,karpov2016raman,agrawal2007nonlinear}
\begin{equation}
    \frac{\partial A(T, t)}{\partial T}=-\left(\frac{\kappa}{2} + i \delta \omega\right) A- \frac{ic\beta_2}{2n_o} \frac{\partial^2 A}{\partial t^2}+\sqrt{{\kappa_{\mathrm{ext}} P_{\mathrm{in}}}}+ ig\left|A\right|^2A
    -i g \tau_\mathrm{A} A  \frac{\partial |A|^2}{\partial t}, 
\label{SI:adiabatic-lle}
\end{equation}
where $\tau_A = f_R\int_{-\infty}^{+\infty}\tau h_R(\tau) \mathrm{d}\tau$ is the Raman shock time. For the full-convolution simulation, we use Eq.~\eqref{SI:lle}. In these simulations, we seed a soliton at a fixed pump power and adiabatically scan the detuning to determine the soliton existence boundary. Raman-induced breathing soliton states \cite{wang2018stimulated,chen2018experimental}—which are dynamically unstable—are not included in the soliton existence region.

\section{Wavelength-dependent transmission}
Wavelength-dependent transmission along the measurement path—from the microresonator to the optical spectrum analyzer (OSA)—results in discrepancies between the measured spectra and the real intracavity spectra. The dominant contributions arise from microresonator–waveguide coupling and fiber components in the path (e.g., couplers).

\begin{figure*}[htbp]
    \centering
    \includegraphics[scale=1.0]{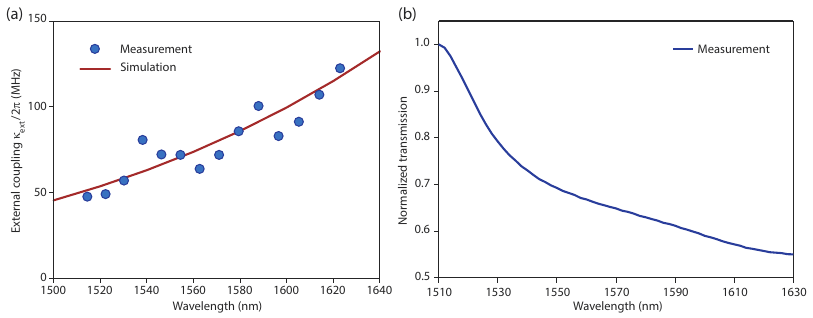}
    \caption{ (a) Measured and simulated wavelength-dependent microresonator–waveguide coupling. (b) Measured transmission of the fiber path from the microresonator to the OSA. 
    } 
    \label{SIFig:coupleDistti}
\end{figure*}

Figure~\ref{SIFig:coupleDistti}(a) presents the measured and simulated external microresonator–waveguide coupling $\kappa_\mathrm{ext}$ as a function of wavelength. $\kappa_\mathrm{ext}$ increases with wavelength, in agreement with the simulation. Figure~\ref{SIFig:coupleDistti}(b) presents the measured transmission $T$ of the fiber path from the microresonator to the OSA. The wavelength dependence of the transmission mainly arises from the wavelength-dependent splitting ratio of the fiber couplers. The transfer relation between the measured spectrum $S_\mathrm{meas}(\omega)$ and the intracavity spectrum $S_\mathrm{intracavity}(\omega)$ is
\begin{equation}
    S_\mathrm{meas}(\omega) \propto T(\omega)\kappa_\mathrm{ext}(\omega) S_\mathrm{intracavity}(\omega).
\end{equation}
Based on this relation, the intracavity optical spectra can be inferred from the measured spectra. 

\section{Self-steepening effect}
In this section, we show that the self-steepening effect is negligible for the microresonator solitons accessible in our experiments. Including self-steepening, the Lugiato–Lefever equation is given by~\cite{agrawal2007nonlinear,bao2015soliton}:
\begin{equation}
    \frac{\partial A(T, t)}{\partial T}=-\left(\frac{\kappa}{2} + i \delta \omega\right) A- \frac{ic\beta_2}{2n_o} \frac{\partial^2 A}{\partial t^2}+\sqrt{{\kappa_{\mathrm{ext}} P_{\mathrm{in}}}}+i g\left(1+\frac{i}{\omega_o}\frac{\partial}{\partial t}\right) \left[   \left(R(t)\ast\left|A(t)\right|^2 \right)A\right].
\label{SI:lle-ss}
\end{equation}
The physical quantities involved in Eq.\eqref{SI:lle-ss} are defined in the main text. The first-order derivative on the right-hand side is responsible for self-steepening.

\begin{figure*}[htbp]
    \centering
    \includegraphics[scale=1.0]{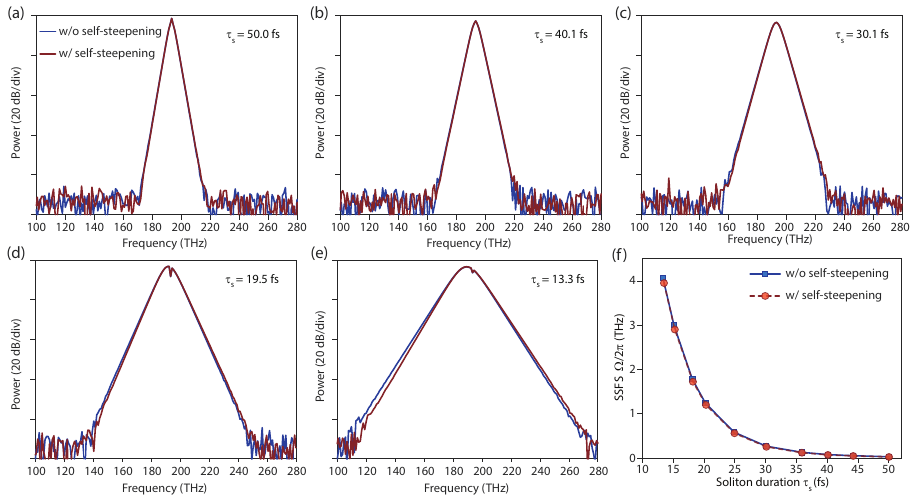}
    \caption{ (a-e) Simulated soliton spectra with self-steepening (red) and without (blue). As the soliton duration decreases, self-steepening becomes more pronounced, yielding asymmetric spectra. (f) Simulated SSFS versus soliton duration $\tau_s$ with self-steepening (orange) and without (blue). The simulations use the same parameters as the Si$_3$N$_4$ microresonator in the main text. Lorentzian Raman response [Eq.~(7) in the main text] is used. 
    } 
    \label{SIFig:selfsteep}
\end{figure*}

Figure~\ref{SIFig:selfsteep}(a–e) shows simulated soliton spectra with self-steepening (red) and without (blue). As the soliton duration decreases, self-steepening becomes more pronounced, yielding asymmetric spectra. Figure.~\ref{SIFig:selfsteep}(f) compares the simulated SSFS as a function of soliton duration with and without self-steepening. In our experiments, microresonator solitons as short as $\tau_s = 13.6~\mathrm{fs}$ are achieved. Simulations indicate that, within the experimentally accessible soliton-duration range, the self-steepening effect can be neglected.

\medskip
\bibliography{ref.bib}